# A 65-nm Reliable 6T CMOS SRAM Cell with Minimum Size Transistors

G. Torrens (Member, IEEE), B. Alorda (Member, IEEE), C. Carmona, D. Malagón-Periánez, J. Segura (Member, IEEE), S.A. Bota (Senior Member, IEEE)

Electronics Systems Group, University of the Balearic Islands, Physics, Palma de Mallorca 07122, Spain
CORRESPONDING AUTHOR: S.A. BOTA (sebastia.bota@uib.es).

This work has been supported by the Spanish Ministry of Science and Innovation under project CICYT-TEC2014-52878-R, financed jointly by FEDER fund.

*Abstract-* **As minimum area SRAM bit-cells are obtained when using cell ratio and pull-up ratio of 1, we analyze the possibility of decreasing the cell ratio from the conventional values comprised between 1.5-2.5 to 1. The impact of this option on area, power, performance and stability is analyzed showing that the most affected parameter is read stability, although this impact can be overcome using some of the read assist circuits proposed in the literature. The main benefits are layout regularity enhancement, with its consequent higher tolerance to variability, cell area reduction by 25% (with respect to a cell having a cell ratio of 2), leakage current improvement by a 35 %, as well as energy dissipation reduction and a soft error rate per bit improvement of around 30 %.**

*Index Terms_-* **SRAM, CMOS embedded SRAM, Circuit reliability, Vt variability, 6T-SRAM.**

## I. INTRODUCTION

The aggressive scaling of CMOS IC technologies has enabled the manufacture of faster and smaller circuits which in the case of SRAM memories has resulted in larger storage capacity and shorter access-time. A significant percentage of the total die area of current System-on-Chip (SoC) is dedicated to memory blocks. One consequence of this fact is that embedded SRAM yield dominates the overall SoC yield [1]. However, with small device dimensions, aggressive design rules, and the ever-increasing demand for on-chip cache capacity, the traditional 6 transistors cell (6T-cell) is now particularly sensitive to device variations and more prone to functional failures than before [2].

SRAM cell size continues to shrink by half each generation [3]. Despite scaling, SRAM cells must be stable during read operations and writeable during write events. The most important barrier to achieving high-yield and small SRAM cells is the large threshold voltage ($V_t$) variability [4], implying that design tradeoffs are generally adapted to accommodate power, performance, and area restrictions.

Cell stability during read operations is enhanced by strengthening the internal latch inverters and weakening the access transistors, while the opposite is desired for cell write-ability: a weak inverter and strong access transistors must be chosen such that read stability and write-ability are both within reasonable levels. This imposes conflicting constraints on 6T-cell transistor sizes [5]. In six transistors bit-cells, acceptable values for both write and read margins have been traditionally obtained by accurate transistor sizing that, as will be shown, impact the bit-cell area. This delicate balance can be severely affected by process variations that dramatically degrade stability and write margins, especially in scaled technologies [5]-[6].

Alternative bit-cell structures have been proposed to overcome the limitations of 6T-cells by separating read and write paths and enabling each operation to be optimized individually at the cost of additional transistors, leading to 8T [7], 9T [8] and 10T cells [9]. Among other proposals that generally imply considerable tradeoffs in floorplanning [10] or cell size [11], the 8T-cell allows reusing traditional SRAM and register file design techniques. Therefore, it is being adopted as an alternative to the conventional 6T-cell in industrial designs [12],[13], specially when a wide range of supply voltages is required to achieve high performance during normal mode while minimizing power consumption, as they allow operation in low voltage mode [7],[14]. Unfortunately, 8T-cells are unable to avoid the problem of half-selected cells (cells that share word-line selection, but are neither written to nor read out during write or read operations). In addition, their read speed can be compromised by the use of unipolar sense amplifiers instead of differential ones, as in 6T-cells. As a complementary alternative, cell stability can also be improved using read and write assist techniques like the ones reported in [15],[16].

Based on these constraints it is usually assumed by default that the pull-down transistors of the internal latches of 6T-cells, must be wider than the access transistors, while the access transistors should be not too small compared with the pull-up transistors (usually they are equal sized).

Therefore, the total cell area is dominated by these two constraints. In this paper we investigate the possibility of using minimum size transistors to achieve a higher bit-cell density compared to both conventional 6T and 8T, while maintaining stability margins within acceptable values. This option is easily extensible to SRAM devices implemented in emerging technologies like FinFET, where device sizes are constrained to be integer numbers of fins, particularly when the dual gate control feature of FinFETs can be used [17],[18]. Our study comprises electrical simulations as well as experimental measurements from SRAM devices fabricated in a 65 nm CMOS commercial technology biased at $V_{DD}$=1.2 V. To obtain realistic simulation data, results have been obtained using a complete column and peripheral circuitry. Given its relevance, the results related to cell stability have been also corroborated with experimental measurements.

## II. SRAM : DESIGN TRADEOFFS

A six transistor memory cell is formed by two access transistors controlled by the word-line (WL) connecting bit-lines $BL_A$ and $BL_B$ with the internal nodes A and B, and two back-to-back connected inverters forming a latch (Fig 1).

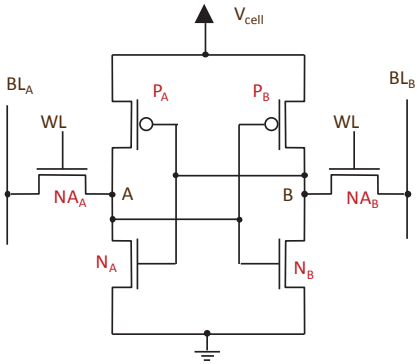

Fig 1.Six transistors memory cell.

### A. Transistor sizing

In 6T-cells, transistor widths must be carefully selected to guarantee cell stability during write and read operations. $W_n$ denotes the width of the pull-down transistors $N_A$ and $N_B$, $W_p$ corresponds to the width of pull-up transistors $P_A$ and $P_B$, $W_{acc}$ is the width of the access transistors $NA_A$ and $NA_B$. Transistors $N_A$ and $P_A$ form the inverter $INV_A$, while $N_B$ and $P_B$ form $INV_B$. In a read operation, the pre-charged bit-lines disturbs the "0" storage node, through the access transistor, by pulling it up. To guarantee a non-destructive read, the cell ratio CR (also known as β ratio), defined as CR=$W_n$/$W_{acc}$, is usually comprised between 1.5 and 2.5 [19]. A 6T-cell having a proper CR to guarantee cell stability during read access is referred in this work as conventional cell or 6T-CC to distinguish it from a 6T-cell with minimum size transistors or 6T-MSC (with CR=1).

A failure to write occurs when the pass transistor is not strong enough to overcome the cell pull-up pMOS transistor. The requirement to perform a write operation is typically met by setting an adequate pull-up ratio (PR= $W_p$/$W_{acc}$) usually lower than 3. To minimize cell area, the size of the pull-up and pass transistors are typically chosen to be minimal (PR=1) [19]. Obviously, the 6T-MSC cell also has PR=1.

### B. Cell Area and Layout

The layout of an SRAM cell determines the array area density and is a key parameter to manufacturing yield of SoCs containing large arrays. In nanometer technologies, the schematic cell of Fig. 1 is implemented using the streamlined layout shown in Fig 2. The uniform orientation of all cell transistors provides better pattern reproducibility and transistor matching [20].

As shown in Fig. 2, bit-cell area is reduced using minimum size transistors; Fig. 3 shows the relative area increase with CR in a cell having PR=1.

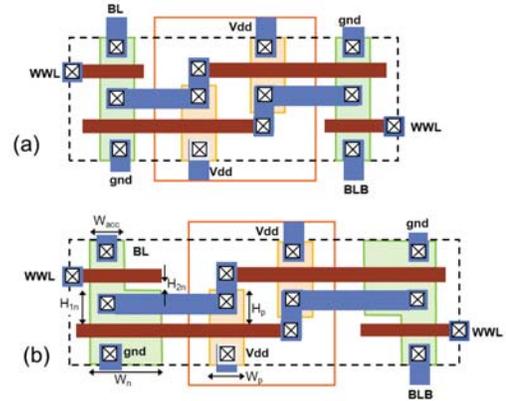

Fig. 2. (a) Layout of 6T-cell with minimum size transistors (b) 6T-cell with CR=2.

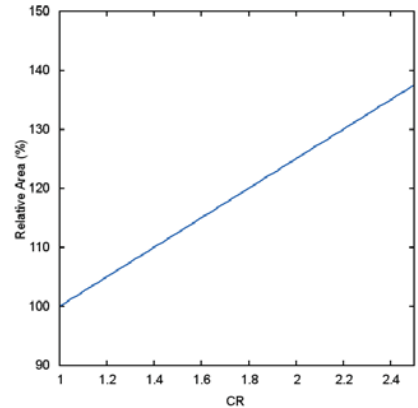

Fig. 3 Relative cell area vs. CR (PR=1).

When considering design for manufacturing (DFM), bit-cell layouts having simple shapes are desirable. It is interesting to point out a 6T-MSC additional advantage provided by its layout, it is much less sensitive to misalignments, given the absence of bends in the n-diffusion [21]. Cell regularity is enhanced and a potential source of variability is removed, which can partially compensate the $V_t$ variation increase due to dopant fluctuation that is inversely proportional to the square root of channel area [22].

### III. CELL STABILITY

SRAM cell data retention, both in standby mode and during a read access is an important functional constraint in advanced technologies. As suggested in the previous section, assuring data retention is the main reason to increase CR instead of keeping all the transistors at its minimum sizes. In any case, when using 6T-MSC cells, the required levels of stability in each specific application must be met.

#### A. Cell Writeability

Write operation can be regarded as a transition from one equilibrium state to the other. The state of the cell will change or not depending on transistor parameters and the magnitude and duration of the perturbation induced during the writing process. Assuming that the cell is initially at state $S_1$ ($V_A$=1 and $V_B$=0), to write the complementary state $S_0$ ($V_A$=0 and $V_B$=1) in the cell, $BL_A$ voltage is decreased to '0', $BL_B$ voltage is raised to '1' and the access transistors $NA_A$ and $NA_B$ are driven ON. Node A being at logic '1' is connected to $BL_A$ at logic '0' and node B at logic '0' to $BL_B$ at logic '1', and, as a result $V_A$ decreases and $V_B$ increases. If $V_A$ is pulled down below the trip point of inverter $INV_B$, the inherent positive feedback of the cell comes into play, $P_B$ becomes stronger than $N_B$ and increases even more $V_B$, which is also the gate voltage of $N_A$, consequently, $N_A$ starts to conduct and further pulls down $V_A$. Eventually $V_B$ rises to $V_{DD}$ and $V_A$ falls to 0, thus flipping the logic states of the cell. Usually PR=1 is sufficient to ensure proper cell writing.

Traditionally, cell writability has been obtained computing the Write Noise Margin (WNM) [23]. Graphically, WNM is illustrated in Fig. 4, being the length of the side of the minimum square spanning the static transfer characteristics of inverter A (measured under $V_{WLA}$ = $V_{DD}$, $V_{BLA}$ = 0 V) and inverter B (measured under $V_{WLB}$ = $V_{DD}$, $V_{BLB}$ = $V_{DD}$). The dependence of WNM on CR is shown in Fig. 5 that also represents WLVM, an alternative parameter used to monitor cell writeability. WLVM is defined as the maximum allowed word-line voltage drop during write operations [24], having the advantage of being easier to measure experimentally. In both cases, it can be noticed the low impact of CR on this parameter, and in any case the best result corresponds to CR=1.

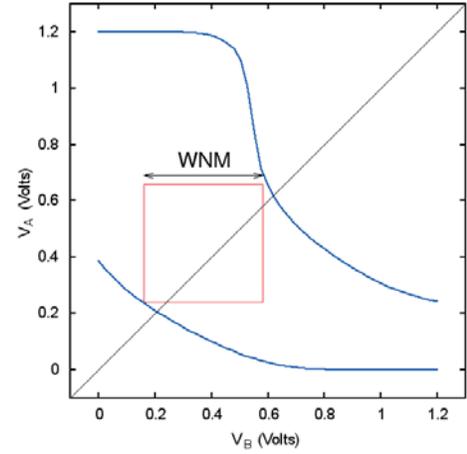

Fig. 4. Graphical definition of WNM.

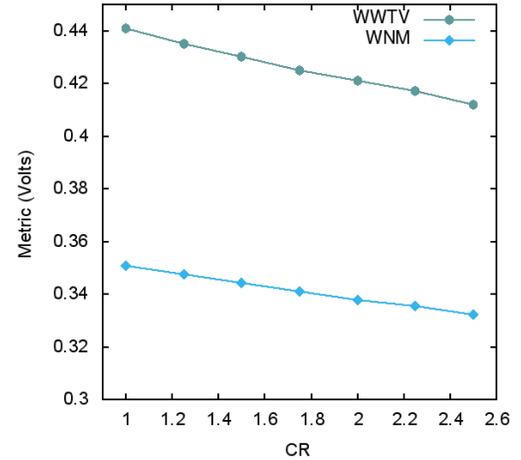

Fig. 5 WNM and WLVM vs. CR (PR=1).

#### B. Read Stability

In a read operation (usually performed with both bit-lines precharged to '1'), the access transistor disturbs the "0" storage node by pulling it up, assuming the cell in $S_1$, the voltage at node B increases to a positive value $V_{READ}$. If $V_{READ}$ becomes higher than the trip point $V_{TRIP}$ of the inverter $INV_A$, then the cell flips while being read. As suggested in [25] $V_{TRIP}$ can be approximated by the inverter $INV_A$ switching voltage.

The most obvious way to maintain a low value for $V_{READ}$ is to enhance the pull-down operation of $N_A$ and $N_B$ by increasing CR. For this reason CR is usually comprised between 1.5 and 2.5 in commercial designs [19]. In Fig 6, we represent the voltage transients of the internal cell nodes for different values of CR during a read access. The corresponding trip-point value is indicated with dashed lines.

A parameter typically used to quantify stability during read is the read static noise margin (RSNM) –defined as the maximum DC voltage noise that can be tolerated by an SRAM cell without losing data during a read operation–. The impact of CR on RSNM is clear when observing the simulation shown in Fig. 7, 6T-MSC cell has a RSNM a 25% lower than the 6T-CC cell.

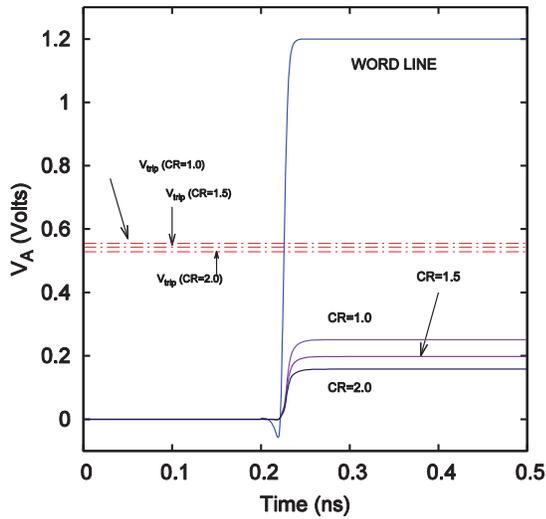

Fig. 6 Voltage transients during read access for CR=1, CR=2, CR=2.5 (PR=1).

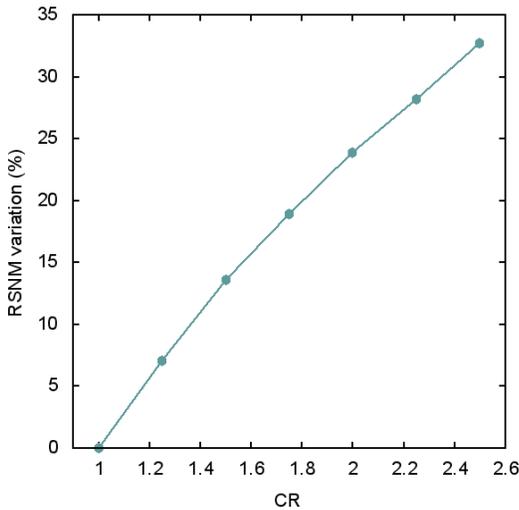

Fig. 7 RSNM vs. CR (PR=1).

Results in Fig. 7 show a remarkable degradation on 6T-MSC read stability, specially when introducing the effects of variability on $V_{READ}$ and $V_{TRIP}$ that discourages its use (we observed an standard deviation of 30 mV in $V_{TRIP}$ and of 34 mV in $V_{READ}$ from Monte Carlo simulation).

A major drawback of the RSNM metric is the inability of being measured in dense SRAM arrays, since the internal storage nodes need to be accessed. The supply read retention voltage (SRRV) [26], is an alternative metric used to account for cell stability during read operations. SRRV is defined as the difference between $V_{DD}$ and the lowest cell supply voltage for data retention in a read cycle while both BL and BLN and WL level are driven by the operating voltage $V_{DD}$. SRRV represents the maximum tolerable DC noise voltage at the bitcell supply before causing the destructive read operation. One of the SRRV advantages is that it is usually easier to measure experimentally than RSNM. It has been reported that SRRV and RSNM cumulative density functions showed an equivalent estimated read failure probability [27].

However, there are alternative ways to reduce the strength of access transistor to avoid read failures while keeping CR=1. In multi-threshold-voltage technologies, it is possible to improve RSNM by selecting the threshold voltages for each transistor in the bitcell [28]. A second alternative is the use of read assist circuits based on word-line voltage modulation. This second option does not involve a permanent design setting and can be applied and adjusted according to the needs of each operation scenario. The read assist circuit improves SNM by slightly lowering the word line voltage. Fig 8 shows that a 10% reduction in word-line voltage in a 6T-MSC during read operation leads to the same RSNM than a 6T-CC at nominal voltage. Such modulation can be achieved by connecting a pull-down transistor in each word-line controlled by the read signal or by using level-programmable drivers to generate intermediate WL levels [29][30]. Both techniques are easily implementable with low area overhead (in [29], a word line voltage modulator circuit was implemented in a 128 kB SRAM macro without area overhead).

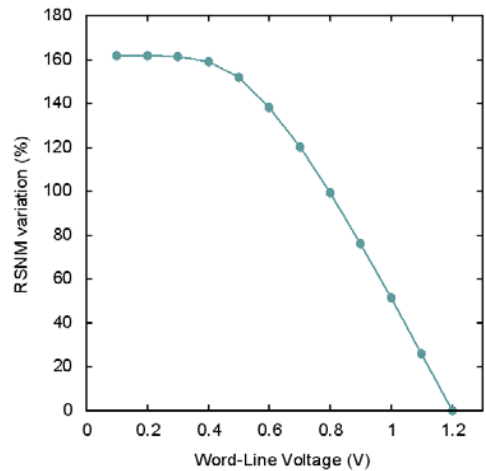

Fig. 8 RSNM vs word line voltage. Note that the value for 0 Volts corresponds to the Static Noise Margin in hold mode.

## C. Soft Errors

Soft errors have risen as a major concern for reliability and dependability of emerging electronic devices [31], specially due to their impact on registers and memories. Neutrons induced by cosmic radiation or alpha particles from packaging materials can flip a memory cell even at sea-level applications. The critical charge, $Q_{crit}$, defined as the minimum amount of charge that a sensitive cell node must collect to induce a change of the cell logic state is reduced because of the decrease of supply voltage and aggressive technology scaling [32]. Therefore, memory devices become sensitive to the disturbances produced by less energetic particles. From electrical simulations, we obtained a 30% higher $Q_{crit}$ for the 6T-CC respect to 6T-MSC for a 65nm CMOS Technology. This result can be understood by noticing the larger internal node capacitance of the 6T-CC respect to the 6T-MSC.

Other studies have shown that $Q_{crit}$ alone is not enough to quantify the device soft error rate [33],[34], since the sensitivity to radiation is also influenced by the charge collection efficiency and cross-section of sensitive nodes. Considering that collection efficiency of nMOS devices is higher than pMOS ones, and the fact that the most sensitive area to soft errors is restricted to the drain region of an off-state transistor, experimental results presented in [33] showed that, despite its higher critical charge, the soft error rate due to alpha radiation in an 6T-CC cell was 40 % worse than in 6T-MSC. Note that final users are interested in reducing soft error rate despite the critical charge is increased or not. The relationship between SER and critical charge is given by [35].

$$SER = \Phi (A_{diff,n} \, exp(-\beta_e Q_{crit,e}) + A_{diff,p} \, exp(-\beta_h Q_{crit,h})) \quad (1)$$

where $\Phi$, denotes the particle flux, $A_{diff,n}$ and $A_{diff,p}$ are the nMOS and pMOS sensitive drain areas. $\beta_e$ and $\beta_h$ describe electron and hole collection density functions. $Q_{crit,e}$ and $Q_{crit,h}$ denote the magnitude of the critical charge for upsets due to the collection of electrons and holes. Critical charge can be related to transistor widths by linear relations as suggested from electrical simulation results [33]:

$$Q_{crit,e} = a_e + b_e W_n + c_e W_p \quad (2)$$

$$Q_{crit,h} = a_h + b_h W_n + c_h W_p \quad (3)$$

parameters $\beta_e, \beta_h, a_e, b_e, c_e, a_h, b_h$ and $c_h$, obtained from fitting of experimental results performed with alpha irradiation of a 65 nm SRAM are reported in Table I.

Fig. 9 shows the dependence of the resulting normalized SER and critical charge with CR. Both SER and critical charge increase with CR. Note that this means that, when increasing CR, critical charge is improved and SER is worsened, therefore, 6T-MSC exhibits lower soft error rate than 6T-CC.

TABLE I
PARAMETERS USED IN SER COMPUTATION (ALPHA PARTICLES) FOR A 65NM CMOS TECHNOLOGY [33].

| Parameter | electrons | holes |
|---|---|---|
| β | 4.95×10$^{14}$ C$^{-1}$ | 1.26×10$^{15}$ C$^{-1}$ |
| a | 0.45 fC | 0.53 fC |
| b | 3.6 fC/μm | 11.3 fC/μm |
| c | 6.5 fC/μm | 2.67 fC/μm |

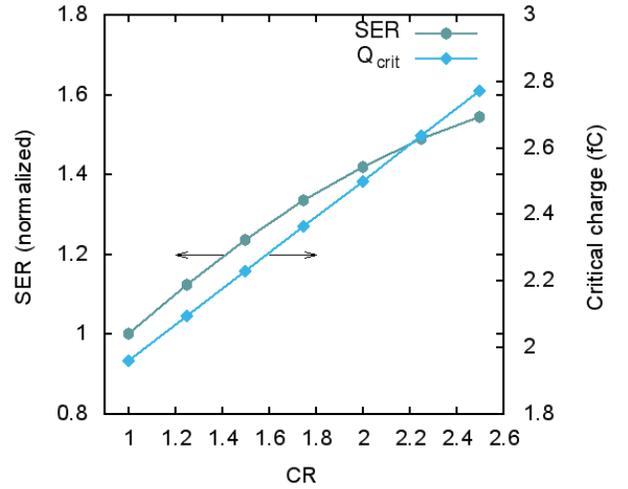

Fig. 9 SER and $Q_{crit}$ vs CR for alpha particles ($Q_{crit}$ obtained from electrical simulation, SER computed using Eqs. (1),(2) and (3))

## IV. POWER

As feature size shrinks, the key component of power consumption will be leakage [36]. Since memories spend a high percentage of their operating time in the hold mode, when considering power consumption it is critical to reduce leakage while the memory is retaining data.

In nanometer CMOS technologies, leakage is dominated by sub-threshold leakage currents, $I_{SUB}$, through the off-state transistors. To clarify the leakage sources in the cell, we will assume the cell is the state $S_1$. Thus $N_A$ and $P_B$ in Fig. 1 are turned off, giving rise to the leakage currents $I_{S1}$ and $I_{S2}$ respectively. $I_{S1}$ and $I_{S2}$ form the cell supply leakage as they drain charge from cell supply grid. Access transistors are off when the word line is not selected (hold mode). Since the bit-lines are pre-charged to $V_{DD}$, $NA_A$ and $NA_B$ both leak. However, since $V_A$ and $V_{BLA}$ are high; resulting in a very small voltage difference between the drain and source of $NA_A$, leakage through $NA_A$ can be reasonably ignored. The leakage through $NA_B$, $I_{S3}$, forms the bit-line leakage.

Each $I_{SUB}$ contribution can be described as [36]:

$$I_{SUB} = \frac{W}{L}\mu V_{th} C_{sth} e^{\frac{V_{GS}-V_t-\eta V_{DS}}{nV_{th}}}\left(1-e^{\frac{-V_{DS}}{V_{th}}}\right) \quad (4)$$

where $V_t$ is the threshold voltage, $V_{th}=qT/k$ is the thermal voltage, µ is the carrier mobility, η is the drain induced barrier lowering, L is the transistor length, W is the width, $V_{DS}$ is the drain to source voltage, $V_{GS}$ is the gate to source voltage, n is the slope shape factor and $C_{sth}$ is the summation of the depletion region capacitance and the interface trap capacitance, both per unit area of the MOS gate.

Fig. 10 represents the total leakage current vs. CR computed at room temperature and nominal voltage for PR=1. We notice a significant leakage reduction, beyond 35% of the 6T-MSC cell respect to the 6T-CC due to the reduction of the $I_{SUB1}$ component.

Fig. 10 also shows the energy consumption related to cell logic state change in a write operation. Again, we report a better figure for the 6T-MSC due to their lower internal node capacitance.

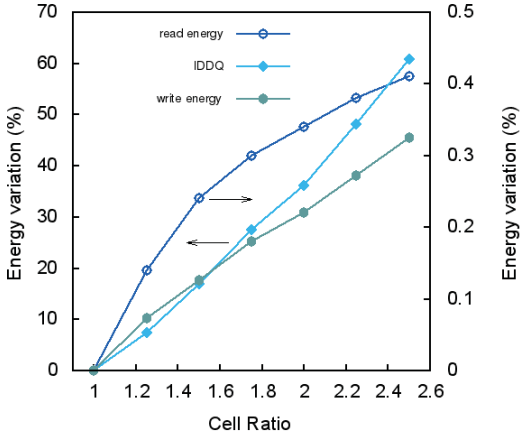

Fig. 10 Power consumption and IDDQ

## V. PERFORMANCE

Although issues associated to read/write timing are not as critical as stability considerations for the SRAM functionality, they may impact overall system performance.

Access transistor ON-current has a dominant influence on the operating speed of the SRAM, so that care should be taken regarding the on-current of the access transistors due to WL modulation during read. Read performance also depends on the read current through the pull-down transistor, read bit-line leakage current, read bit-line capacitance and sensing delay of the sense amplifier. Intrinsic read delay was measured as the time interval from the 50% point of the read wordline signal low-to-high transition until one of the bit-line voltages dropped to the sense amplifier threshold; measurements were performed having 256 memory cells attached to the bit-line and assuming than the read cell was in state $S_1$. Simulation results are illustrated in Fig. 11. Changing the cell ratio from 2.0 to 1.0 will degrade the average read time from 190 ps to 237 ps, (+25%) due to the decrease of the read current through $N_2$. This delay is also affected by the application of a read assist technique, the intrinsic read delay rises to 272 ps when the word-line control voltage is lowered from 1.2 to 1V to increase RSNM.

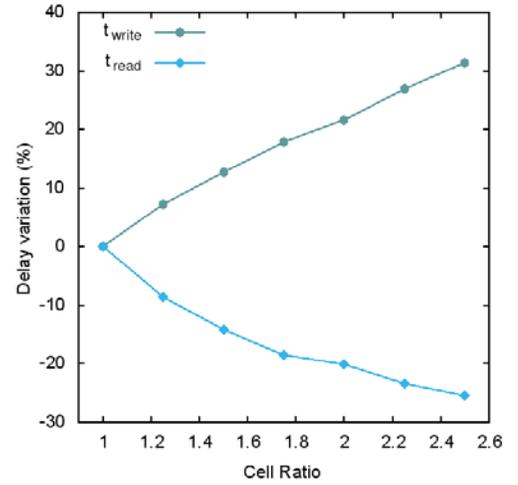

Fig. 11 Timing vs. CR

WL level is not lowered during write. In the transient simulation of intrinsic write performance, full swing voltages were applied on the bit-lines –the intrinsic write delay was measured as the time interval from the 50% point of the write wordline signal low-to-high transition until one of the data storage nodes is pulled up to $V_{DD}/2$ (from an initial voltage of $V_{GND}$) during a write operation. Simulation results are illustrated in Fig. 11. Changing the cell ratio from 2.0 to 1.0 improves the intrinsic write time from 30.0 to 24.0 ps. Write time is mainly affected by the pull-up ratio [19], as this ratio is maintained to 1, write time increases with CR due to the increase of the internal node capacitances related to transistor width increase. In any case, write operations will not represent the limiting factor on overall system performance since the intrinsic write time is more than one order of magnitude faster than reading due to the much smaller load capacitance involved during the write procedure.

Usually the overall read/write delay is much higher than the intrinsic read/write delay, and the benefits of area improvements related to the 6T-MSC outweigh the detrimental impact on overall speed loss. Improvements in both performance and area may not be simultaneously possible. As a conclusion, we may shift the emphasis into improving system functionality by increasing device density, without an excessive loss in performance (a similar performance loss is also produced when using a 8T cell instead a conventional 6T one [34]).

## VI. EXPERIMENTAL RESULTS

A 16 kb SRAM was fabricated on a 65 nm CMOS technology with a nominal supply voltage of 1.2 V. It consists of one SRAM cell array of 6T-MSC cells with 256 rows and 64 columns, with independent bit-line, word-line and bitcell voltages. The cell array was divided into eight sub-blocks composed of eight columns, generating one bit per sub-block (an IO per sub-block). The prototype features independent control of $V_{cell}$, WL and BL levels. $V_{WL}$ modulation has been achieved by connecting the power supply rail of the last row-decoder stages to a dedicated supply I/O. This solution was chosen to facilitate prototype testing, although it can be substituted by a read-assist module like in [29].

March tests were applied and showed that all bitcells were functional when working at nominal conditions ($V_{DD}$=1.2 V, room temperature, at the maximum speed allowed by our test set-up). To account for the impact of process variation on writeability we measured the WLVM obtaining the Gaussian distribution shown in Fig. 12. Its mean value was 405 mV with a standard deviation of 42 mV. As expected, the large local $V_t$ variability degraded the SRAM write margin. In any case, even the cells exhibiting a higher resistance to be written, had a sufficient margin to avoid writeability malfunction.

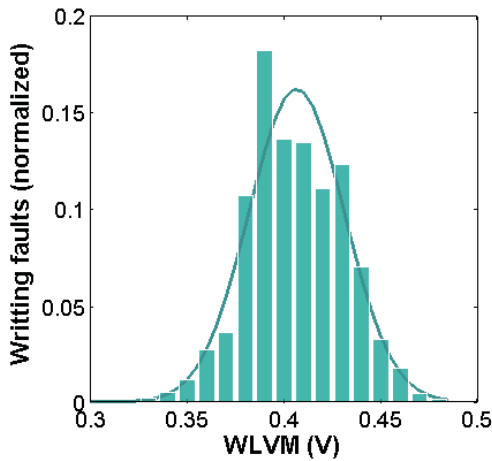

Fig. 12 Measured WLVM histogram for the 6T-MSC cell.

The SRRV was also measured to characterize cell behavior during read (Fig. 13). Under nominal conditions, we obtained a SRRV mean value of 309.8 mV with standard deviation of 51.6 mV. Assuming Gaussian distributions, a good approach for the weaker cells with SRRV values close to zero, the probability of having a cell with SRRV lower than zero (a cell that will flip during read) is $1.00\times10^{-9}$. This value is increased to $1.19\times10^{-4}$ in the worst-case scenario that $V_{cell}$ is reduced by 10%. SRRV increases to 451.4 mV with standard deviation of 47.8 mV when word-line voltage is decreased from 1.2 V to 1.0 V during read access, the probability of having a cell with SRRV lower than zero is $1.893\times10^{-21}$ for $V_{cell} = V_{dd}$ and $2.1\times10^{-12}$ for $V_{cell} = 0.9\ V_{dd}$.

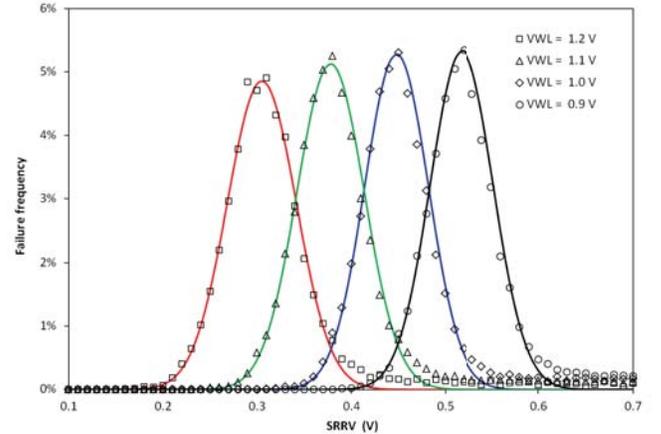

Fig. 13. Experimental distribution of SRRV in the 6T-MSC cell SRAM.

6T-MSC has a stable operating margin, employing some assist circuitry to improve the read operating margin in the presence of variability. Finally, note that the application of word-line voltage modulation during write will also improve the stability of half-selected cells in both 6T and 8T architectures. Half-selected cells can be regarded as cells in read mode in terms of stability. According to the obtained experimental results, the application of word-line-voltage modulation technique will be feasible during read and write operations. Therefore, it will be possible to choose $V_{WL}$ as a trade-off to improve the stability of both the read-out and half-selected cells without causing an excessive impact on cell writeability.

## VII. CONCLUSIONS

We review the advantages of using minimum size transistors in 6T SRAM cells considering all the important aspects of such circuit. The most obvious advantage is related to cost as a consequence of cell area reduction. Minimum size transistors also imply designs having cell ratio = 1 and pull-up ratio = 1, allowing for the adoption of a DFM friendly physical layout using straight diffusion and poly-Si layers. Minimum size transistors also lead to less leakage currents and lower energy involved in the process of changing the logic state of the cell due to smaller internal capacitances. Cell stability was also analyzed: while writeability is maintained, read stability is negatively affected (RSNM is reduced a 25% respect to a cell with

CR=2). However, we show that this potential drawback can be neglected by decreasing the access transistor strength when reducing the word-line level during cell reading. It is also interesting to state the improvement of Soft Error Rate due to radiation given their smaller cell cross-section (although minimum-size cells have lower critical charge).

Performance is affected during the read process; the intrinsic cell delay required to unbalancing the bit-lines depends on the access transistors strength. If such strength is reduced by the read assist circuitry, read time can increase up to 25 % when reducing the cell ratio from 2 to 1. However this percentage is reduced when considering the overall delay (not only the one related with the bitcell).

An SRAM with cells having CR=1 and PR=1 was fabricated on a 65 nm CMOS technology to account for the impact of variability in the previous simulation results, specifically the issues related to stability. Our results confirmed the presence of an adequate noise margin during writing, but also the degradation of SRRV, which can cause problems specifically in weak cells. In this sense, we also verified experimentally, the possibility to solve this problem by adjusting the word-line level during read.

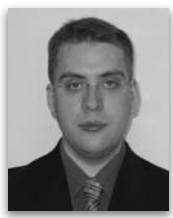
**GABRIEL TORRENS** received the M.S. degree in engineering from the Polytechnic University of Catalonia, Barcelona, Spain, in 2004 and the Ph.D. degree in electronics engineering from the University of the Balearic Islands, Palma de Mallorca, Spain, in 2012. He is currently a part-time Adjunct Professor and a Researcher with the Electronic Systems Group, University of the Balearic Islands. His research interests include VLSI design, CMOS reliability, and radiation-induced phenomena.

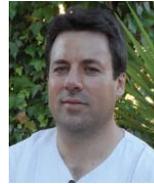
**BARTOMEU ALORDA** received the B.S. degree from the University of Balearic Islands, in 1995, and the M.S. in Telecomunication Engineering from Politechnic University of Catalonia in 2000. The Ph.D. degree in Physics was received from the University of Balearic Islands in 2005. He is currently an associated professor at the Illes Balears University. The main interests are related to memory reliability issues and the exploration of new devices to improve memory capabilities.

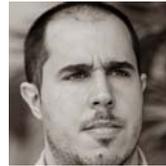
**CRISTIAN CARMONA** received the master's degree in electronic engineering from the University of Balearic Islands in 2014. He is currently a Pre-Doctoral Researcher with the Department of Physics of the University of Balearic Islands. His primary research interest is the analysis and implementation of metrics to improve stability in SRAM, as well as applied energy efficiency methodologies**.**

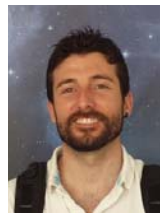
**DANIEL MALAGÓN-PERIÁNEZ** received the B.Sc degree in Physics in 2011 from the University of Seville and the M.Sc. degree in microelectronics, design and applications of micro/nano systems in 2014 from the University of Seville. He is currently a PhD-student with the University of Balearic Islands, Spain. His fields of interest are radiation effects in micro/nano electronics devices and learning algorithms to apply in data analysis.

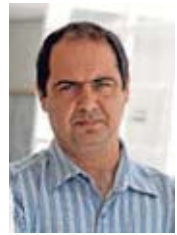
**JAUME SEGURA** received the M.S. degree from the University of the Balearic Islands (UIB), Palma de Mallorca, Spain, in 1989, and the Ph.D. degree from the Polytechnic University of Catalonia, Catalonia, Spain, in 1992. He has been a Full Professor at UIB since 2007. His current research interests include device and circuit modeling and VLSI design and test.

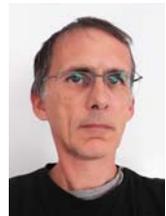
**SEBASTIÀ BOTA** (SM'15) received the M.S. degree in physics and the Ph.D. degree in microelectronics from the University of Barcelona, Barcelona, Spain, in June 1987 and June 1992, respectively. From 1988 to 2002, he was with the Department of Electronics, University of Barcelona, where he became an Associate Professor in electrical engineering in 1995. Since 2002, he has been an Associate Professor with the Electronic Technology Group, Physics Department, University of the Balearic Islands, Palma, Spain. His research interests include VLSI design for nanoscale silicon technologies, CMOS reliability, and circuit-level and physical-design-related test issues.